\documentclass[aps,twocolumn,showkeys]{revtex4}
\usepackage{color}
\usepackage{latexsym}
\usepackage{mathrsfs}
\usepackage[]{graphicx}

\definecolor{grey}{rgb}{0.75,0.75,0.75}
\definecolor{orange}{rgb}{1.0,0.5,0.5}
\definecolor{brown}{rgb}{0.5,0.25,0.0}
\definecolor{pink}{rgb}{1.0,0.5,0.5}

\begin{document}

\title{Functional Renormalization Group Flow of Massive Gravity}

\author{Maximiliano Binder}
\author{Iv\'an Schmidt}
\affiliation{Universidad T\'ecnica Fed\'erico Santa Mar\'ia}
\affiliation{Centro Cient\'ifico Tecnol\'ogico de Valparaiso (CCTVal)}

\keywords{Asymptotic Safety, Massive Gravity, Functional Renormalization}

\begin{abstract}

We apply the functional renormalization group equation to a massive Fierz-Pauli action in curved space and find that, even though a massive term is a modification in 
the infrared sector, the mass term modifies the value of the non-gaussian fixed point in the UV sector.We obtained the beta function for the scale dependent mass parameter 
and found that the massive Fierz-Pauli case still seems to be an asymptotically safe theory. 

\end{abstract}

\maketitle

\section{Introduction}

From a phenomenological point of view, General Relativity \cite{Einstein} is one of the best physical theories that we have at the present time. It explains with great accuracy macroscopical phenomena that could not be accounted for
with newtonian mechanics, such as the bending of light by a massive object and the precession of Mercury's perihelion. Problems arise when one tries to give a quantum mechanical 
description of gravity following well known quantization procedures. This quantum theory of gravity presents divergences in the UV sector \cite{t'hooft}, which render it non-renormalizable by the usual perturbative approach, and therefore General Relativity 
is usually considered an effective field theory valid up to some energy scale \cite{Donoghue}. Several attempts have been made in order to obtain a quantum theory of gravity, such as string 
theory, non commutative geometry, loop quantum gravity, etc., and one of these is Asymptotic Safety \cite{WeinbergAsymptotic,Reuter98,Percacci}.
 
Asymptotic Safety is a set of conditions that ensures the good behavior of a theory in the full energy scale, which means that the theory under consideration can be considered a
fundamental theory. This approach has been used extensively in order to understand the UV behavior of gravity, in a way that avoids the non-renormalizability of the quantum theory
of gravity in the perturbative quantum field theory formalism. The Asymptotic Safety conditions are based on the existence of fixed points in the flow of the couplings constants as
functions of the energy scale. It has been used for many gravity Lagrangians, such as Einstein-Hilbert, F(R) type, torsion, etc...  In all these cases, gravity appears 
to be an asymptotically safe theory \cite{HolstReuter,higherderivative,hdreuter,Topological}.

In quantum field theory, the particle mediating the gravitational interaction is a spin 2 massless particle \cite{WeinbergCampos}, which is always attractive and produces a 
long range force. On the other hand, observations show that the universe is accelerating in its expansion \cite{expanding}, and this is in contradiction with the principle of a universal
attractive interaction.  Possible explanations could come from modifications of general relativity in the infrared sector, such as giving a mass to the graviton, resulting in a massive gravity theory \cite{Hinterbichler}. Studying massive gravity is also important in extra-dimensional extensions of the Standard Model of fundamental interactions,  since in these theories the higher Kaluza-Klein modes are usually massive gravitons. 

In this paper we study how the beta functions for the Newton's and cosmological coupling constants get modified if we include a mass term in the Einstein-Hilbert action.
This theory has been proven to have theoretical problems \cite{Vainshtein}, particularly in its ultraviolet completion, and therefore it is important to see whether and how these difficulties are still present when it is considered from the point of view of the renormalization group.
Here we will not enter into details about massive gravity, and for that purpose the reader can see any of the several references \cite{Hinterbichler}. Our main goal in this work is to study how the known flow diagrams, 
and the fixed points of the coupling constants in the Einstein-Hilbert truncation get modified when a mass term is added to the action. The $m\rightarrow 0$ limit is 
not a trivial one already in non-abelian gauge theories and this could be appreciated in the flow diagrams or in the beta functions.

\section{Massive Fierz-Pauli Truncation}

We follow the same procedure as for the Einstein-Hilbert truncation in \cite{Reuter_calculos}, but we include a mass term in the action

\begin{eqnarray}
\Gamma_{k}[g,\bar{g}]&=&2\kappa^{2}Z_{Nk} \int d^{d}x\Big[ \sqrt{g} (-R + 2 \bar{\lambda}_{k}) \nonumber \\
& &-\sqrt{\bar{g}}\frac{1}{2}m^{2}(h_{\mu\nu}h^{\mu\nu}-h^{2})\Big]
\label{lagrangiano}
\end{eqnarray}

Here $g,\bar{g}$ and $h$ are related by $g_{\mu\nu}=\bar{g}_{\mu\nu} + h_{\mu\nu}$ where $\bar{g}$ is a fixed background metric, $h_{\mu\nu}$ is the fluctuating field that enters in the path integral
and $Z_{Nk}$ is related to the running  Newton constant $G_{k}\equiv\bar{G}/Z_{Nk}$ with $\bar{G}$ a fixed constant and $\bar{\lambda}_{k}$ is the dimensionfull cosmological constant. The construction
of the mass term in this cases is the only one possible that do not generates a massive scalar ghost when linearizing (\ref{lagrangiano}). So this action describes a single spin 2 massive graviton. Since we are working without
the presence of a source, $T^{\mu\nu}=0$, we can restore the diffeomorphism invariance using the Stuckelberg mechanism \cite{Stuckelberg} .
The evolution equation for $\Gamma_{k}$ is given by the Wetterich equation \cite{Wetterich}

\begin{widetext}
\begin{eqnarray}
\partial_{t}\Gamma_{k}[\bar{h},\xi,\bar{\xi};\bar{g}] &=& \frac{1}{2}Tr\Bigg[\Big(\Gamma^{(2)}_{k}+R_{k}\Big)^{-1}_{\bar{h}\bar{h}}\Big(\partial_{t}R_{k}\Big)_{\bar{h}\bar{h}}\Bigg] - \frac{1}{2}Tr\Bigg[\Bigg\{\Big(\Gamma^{(2)}_{k}+R_{k}\Big)^{-1}_{\bar{\xi}\xi}-\Big(\Gamma^{(2)}_{k}+R_{k}\Big)^{-1}_{\xi\bar{\xi}}\Bigg\}\Big(\partial_{t}R_{k}\Big)_{\bar{\xi}\xi}\Bigg]  
\label{evolutionequation}
\end{eqnarray}
\end{widetext}
where the second term is for the Fadeev-Popov ghosts ($\bar{\xi}$ and $\xi$), which are included as in Yang-Mills theories and in this case we are working with the harmonic gauge 
$\partial^{\mu}h_{\mu\nu}-\frac{1}{2}\partial_{\nu}h=0$ .

In order to obtain the beta functions for $\bar{\lambda}_{k}$ and $Z_{Nk}$, we consider the evolution equation in the case $g_{\mu\nu}=\bar{g}_{\mu\nu}$ 
(or $h_{\mu\nu}=0$), after performing the derivatives $\Gamma^{(2)}_{k}=\frac{\delta^{2}}{\delta h(x)\delta h(y)} \Gamma_{k}$. The same holds for the ghost part 
with $\bar{\xi}=\xi=0$. Thus the left hand side of the evolution equation results in

\begin{equation}
 \partial_{t}\Gamma_{k}[g,\bar{g}]=2\kappa^{2} \int d^{d}x \sqrt{\bar{g}} \Big[-(\partial_{t}Z_{Nk})R + 2 \partial_{t}(Z_{Nk}\bar{\lambda}_{k})\Big]
\end{equation}

As we can see, taking the limit $h\rightarrow0$ in the LHS of the evolution equation supreses all the information of $\partial_{t} m(k)$, and therefore in this case 
we can consider the mass as a scale dependent coupling or just a parameter with mass dimensions.

The right hand side of (\ref{evolutionequation}) is much more technical to obtain. For details the reader can see \cite{Reuter_calculos}, but the recipe is straight forward: first 
obtain $\Gamma^{(2)}_{k}$ by expanding $g_{\mu\nu}=\bar{g}_{\mu\nu}+h_{\mu\nu}$ and then take the derivative $\frac{\delta}{\delta h(x)\delta h(y)}$. Then, consider 
the fourier transform for the trace  $Tr[(\Gamma^{(2)}+R_{k})^{-1}\partial_{t}R_{k}]=Tr[W(-D^{2})] = \int ds \widetilde{W}(s) Tr[e^{-isD^{2}}]$ and use the Heat kernel 
expansion to evaluate $Tr\Big[e^{-isD^{2}}\Big]$. After performing a Mellin transform we can compare the coefficients of $\int d^{4}x \sqrt{g}$ and $\int d^{4}x \sqrt{g} R$ 
in the LHS and RHS of the evolution equation and obtain a system of two equations (a brief description is given in Appendix \ref{sss}).

One obtains the following set of equations for the cosmological and gravitational coupling constants  $\bar{\lambda}_{k}$ and $Z_{Nk}$:

\begin{widetext}
\begin{eqnarray}
 \partial_{t}(Z_{Nk}\bar{\lambda}_{k})&=& \frac{1}{16\kappa^{2}}(4\pi)^{-\frac{d}{2}}k^{d}\Bigg[d\Big(d+1)\Big\{2Q^{1}_{\frac{d}{2}}-\eta_{N}(k)\widetilde{Q}^{1}_{\frac{d}{2}} \Big\} - 8dQ^{1}_{\frac{d}{2}}(0) +\frac{m^{2}}{k^{2}}\frac{(d-1)(d^{2}-2)}{(d-2)}\Big\{2Q^{2}_{\frac{d}{2}}-\eta_{N}(k)\widetilde{Q}^{2}_{\frac{d}{2}} \Big\}\Bigg] \nonumber \\
\partial_{t}Z_{Nk}&=&-\frac{1}{4\kappa^{2}}(4\pi)^{-\frac{d}{2}}k^{d-2}\Bigg[\frac{d}{12}\Big(d+1\Big)\Big\{2Q^{1}_{\frac{d}{2}-1}-\eta_{N}(k)\widetilde{Q}^{1}_{\frac{d}{2}-1} \Big\} -\frac{d}{2}\Big(d-1\Big)\Big\{2Q^{2}_{\frac{d}{2}}-\eta_{N}(k)\widetilde{Q}^{2}_{\frac{d}{2}} \Big\} \nonumber \\ 
& &-\frac{2}{3}dQ^{1}_{\frac{d}{2}-1}(0)- Q^{2}_{\frac{d}{2}}(0) - \frac{m^{2}}{k^{2}}\frac{d^{2}(d-1)}{24(d-2)}\Big\{2Q^{2}_{\frac{d}{2}-1}-\eta_{N}(k)\widetilde{Q}^{2}_{\frac{d}{2}-1} \Big\}\Bigg]
\label{dimensionfulleqs}
\end{eqnarray}
\end{widetext}

Here $\widetilde{Q}^{m}_{n}$ and $Q^{m}_{n}$ are the ``threshold functions" mentioned in \cite{Reuter98,Reuter_calculos}, which are integrals depending on the form of $W[-D^{2}]$ and the cutoff $R_{k}$. 
In our case we consider the sharp cutoff $ R_{k}(p^{2})=\lim_{\hat{R}\rightarrow\infty}\hat{R}\Theta(1-\frac{p^{2}}{k^{2}})$, so that the integrals that appear can be 
evaluated in analytic form. The equations (\ref{dimensionfulleqs}) are for the dimensionfull couplings $Z_{Nk}$ and $\bar{\lambda}_{k}$, while for the dimensionless couplings 
$\lambda_{k} \equiv k^{-2}\bar{\lambda}_{k}$, $g_{k} \equiv \bar{G}k^{d-2}Z^{-1}_{Nk}$ and $m_{k}\equiv \frac{m}{k}$ one obtain the beta functions 

\begin{widetext}
\begin{eqnarray}
\partial_{t}g_{k}&=& (d-2+\eta_{N}(k))g_{k} \nonumber \\
\partial_{t}\lambda_{k}&=&-(2-\eta_{N})\lambda_{k}-\frac{g_{k}}{\pi}\Bigg[5ln(1-2\lambda_{k}) -2\zeta(3)+\frac{5}{2}\eta_{N} -m^{2}_{k}\frac{21}{8}\Big(-\frac{\eta_{N}}{2}-2ln(1-2\lambda_{k})-4\zeta(3)\Big)\Bigg],
\label{funcionesbeta}
\end{eqnarray}
where $\eta_{N}(k)=-\partial_{t}lnZ_{Nk}$ is the anomalous dimension of the R operator.

\begin{eqnarray}
\eta_{N}&=&-\frac{2g_{k}}{6\pi+5g_{k}}\Bigg[\frac{18}{1-2\lambda_{k}}+5ln(1-2\lambda_{k})-\zeta(2)+6 + m^{2}_{k}\frac{3}{1-2\lambda_{k}}\Bigg]  
\end{eqnarray}
\end{widetext}

As we can see in (\ref{funcionesbeta}), there is no discontinuity in the $m^{2}_{k}\rightarrow0$ limit and one can reproduce the massless results of \cite{Reuter_calculos} taking 
the massless limit of the massive case.  It is important to notice that in (\ref{dimensionfulleqs}) the mass appears as $\frac{m^{2}}{k^{2}}$, so it automatically results as a
dimensionless term in the dimensionless flow equations.

\begin{widetext}
\begin{figure}
\begin{center}
\begin{tabular}{ccc}
	  \includegraphics[width=0.3\textwidth]{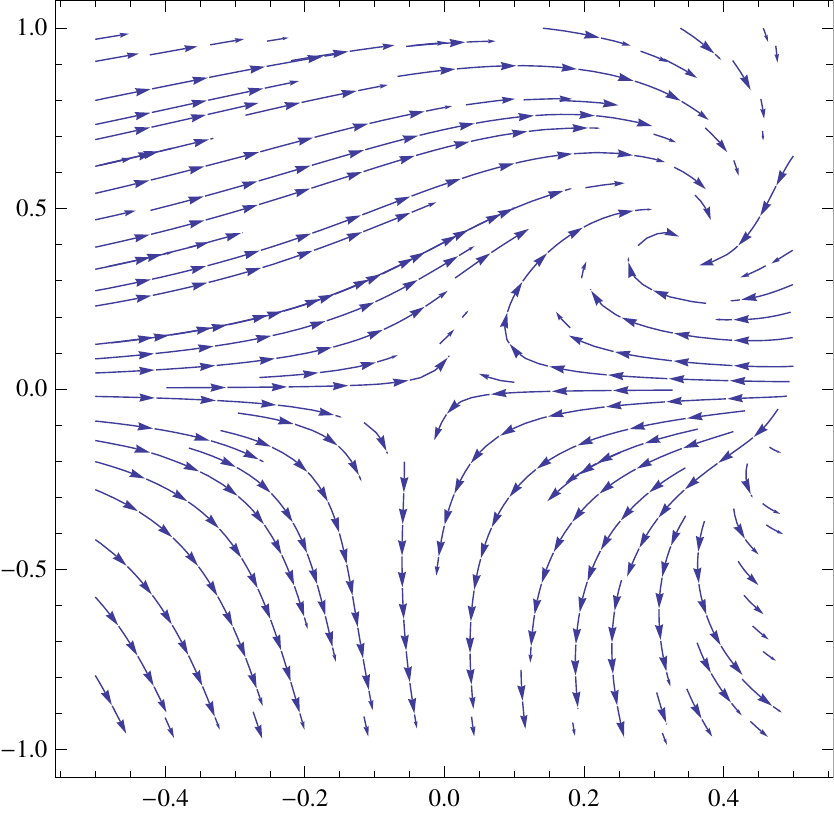}
&	  \includegraphics[width=0.3\textwidth]{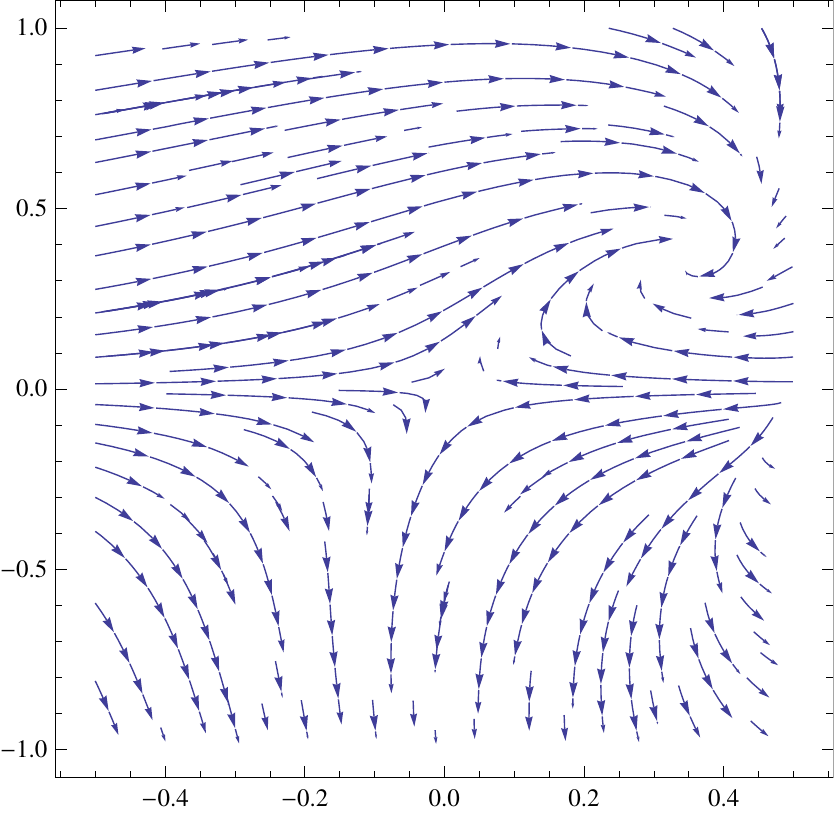}
&	  \includegraphics[width=0.3\textwidth]{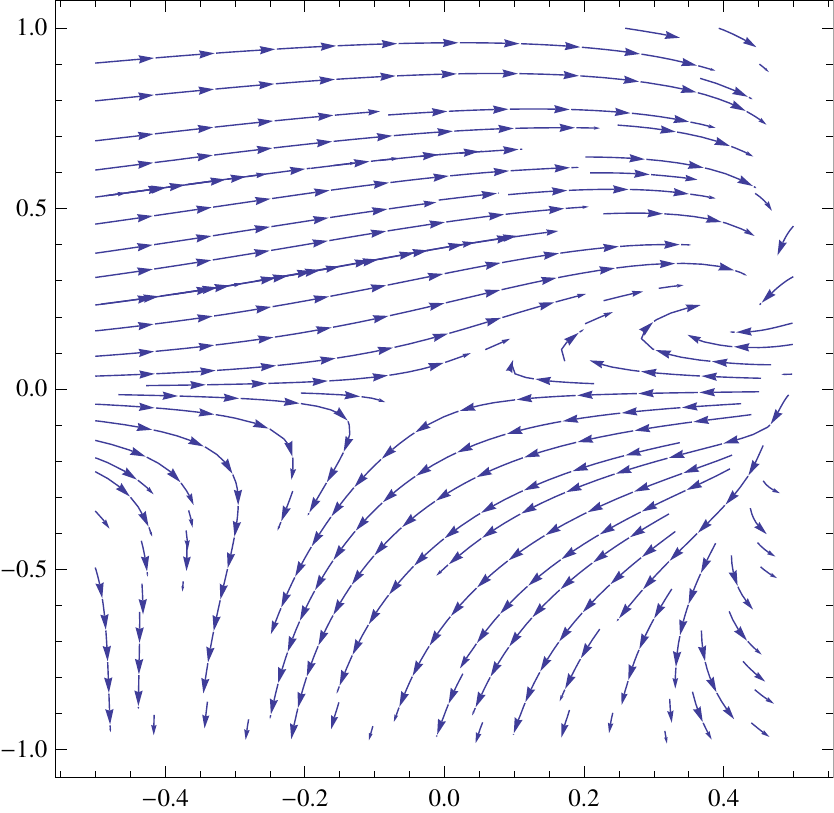}
\end{tabular}
\caption{Flow diagrams for $m^{2}_{k}=0,0.1$ and $0.5$. Here $\lambda_{k}$ ($g_{k}$) is in the x (y) axis. The coordinates for the $m=0$ fixed point are given by $\lambda_{k}=0.3296$ and $g_{k}=0.40266$}
\label{flujodistintasmasas}
\end{center}
\end{figure}
\end{widetext}.

As we can see in Fig.\ref{flujodistintasmasas}, the flow diagrams change continuously as the dimensionless mass $m^{2}_{k}$ grows.

\begin{figure}
	  \includegraphics[width=0.45\textwidth]{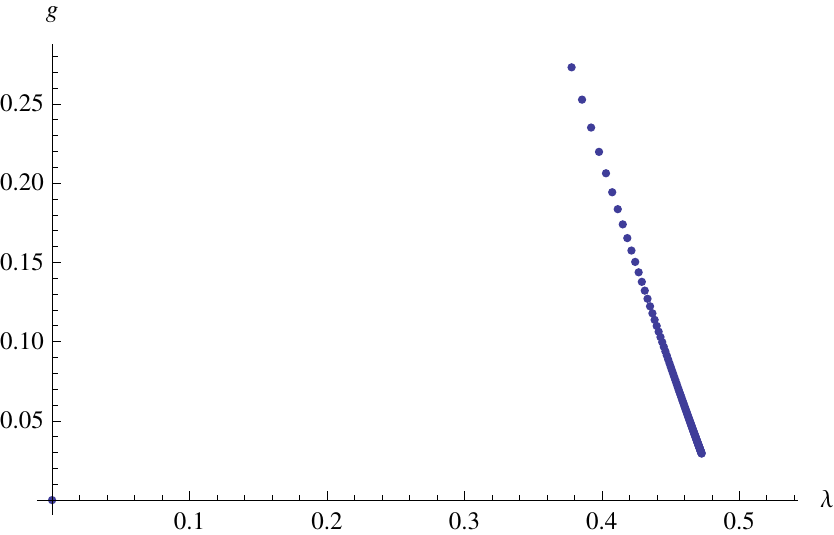}
	  \caption{Coordinates of the non-gaussian fixed point for different values of the dimensionless mass $m^{2}_{k}$.}
	  \label{coordenadaspuntosfijos}
\end{figure}

The non gaussian fixed point gets modified as the dimensionless mass grows (Fig.\ref{coordenadaspuntosfijos}). In this case we have taken values for $m^{2}_{k}$ that are consistent with the interval
$0<\lambda_{k}<0.5$, to avoid problems with the $\ln(1-2\lambda_{k})$ term in (\ref{funcionesbeta}). For this purpose we have chosen values in the range $0\leq m^{2}_{k}\leq6$.

\begin{figure}
	  \includegraphics[width=0.45\textwidth]{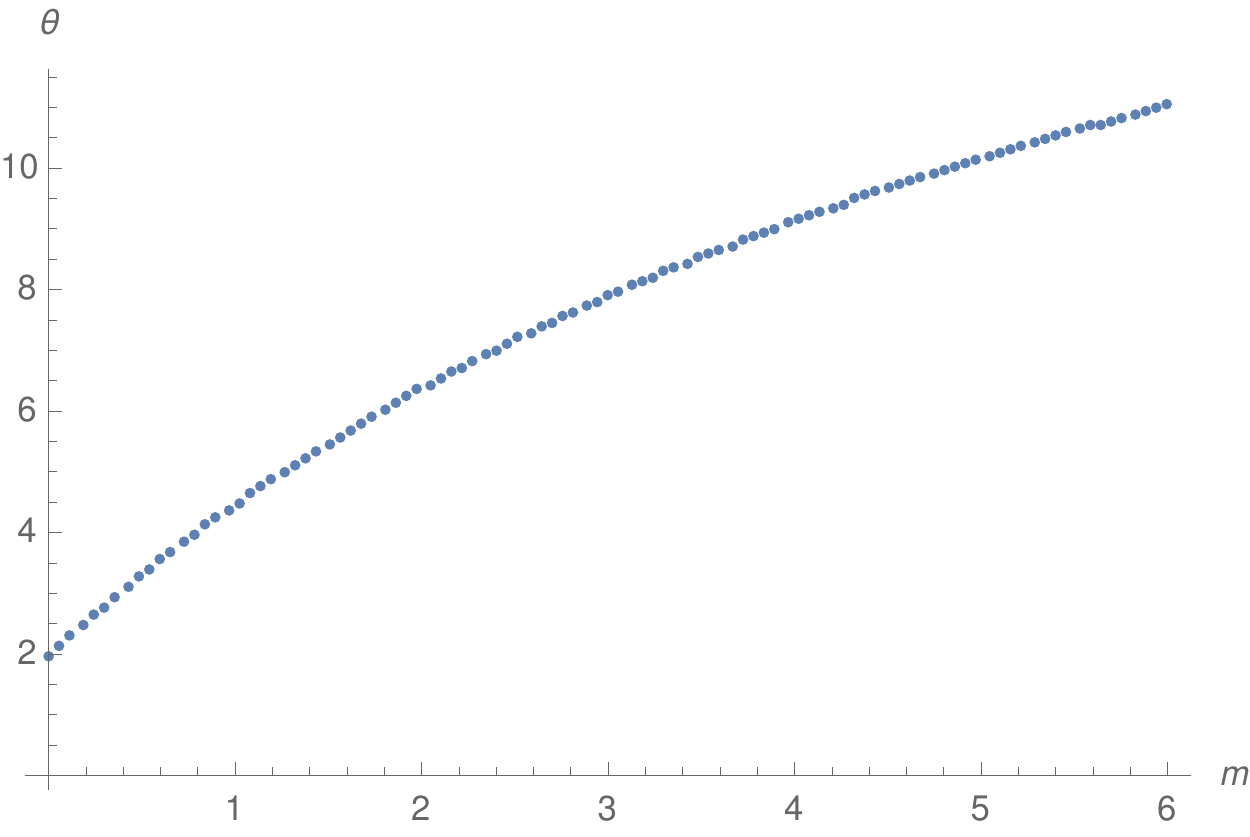}
	  \caption{The real part $\theta$ of the critical exponent is positive and grows with the dimensionless mass $m^{2}_{k}$.}
	  \label{criticalexponents}
\end{figure}

We can also obtain the behavior of the critical exponents depending on the value of the mass parameter (Fig. \ref{criticalexponents}). The values of critical exponents for the $m_{k}=0$ 
case are $\theta_{1}=\theta^{*}_{2}=1.94091 - 3.31065 i$. Therefore, we have two complex conjugate critical exponents whose real part is positive and grows together with the mass parameter, 
ensuring an asymptotically safe behavior around the non-gaussian fixed point for both the massive and for the massless case.

\section{Mass Beta Function}

We can consider the mass term as a scale dependent parameter $m_{k}=m(k)$ and obtain its beta function by taking field derivatives of the Wetterich equation following the procedure described in 
detail in \cite{codello} for the case of non-abelian massive theories (similar applications have been done for the case of gravity \cite{globalflows} but not in a massive gravity context ). 
Since the mass term comes multiplied by a quadratic term in the field $h_{\mu\nu}$, we consider a second order functional derivative of equation (\ref{evolutionequation})
to obtain

\begin{eqnarray}
\partial_{t}\Gamma^ {(2)}_{k} &=& Tr \Big\{ G_{k}\Gamma_{k}^{(3)}G_{k}\Gamma_{k}^{(3)}G_{k}\partial_{t}R_{k}\Big\} \nonumber \\
& & -\frac{1}{2} Tr \Big\{ G_{k}\Gamma_{k}^{(4)}G_{k}\partial_{t}R_{k}\Big\},
\label{wetterich2deriv}
\end{eqnarray} 
where $G_{k}= [\Gamma_{k}^{(2)}+R_{k}]^{-1}$ is the modified propagator and $\Gamma_{k}^{(n)}$ is the n-order functional derivative  
$\frac{\delta^{n}}{\delta h\delta h \cdots}\Gamma_{k}$ with respect to the fluctuating field $h_{\mu\nu}$. So we need the expansion up to fourth order in $\sqrt{g}\rightarrow\sqrt{\bar{g}}(O(h^{0})+O(h^{1})+O(h^{2})+O(h^{3})+O(h^{4}))$ 
and the same for the curvature R with $g_{\mu\nu} \rightarrow \bar{g}_{\mu\nu}+h_{\mu\nu}$. After taking the field derivatives and contracting the indices  we obtain for the left hand side of 
(\ref{wetterich2deriv})

\begin{eqnarray}
\partial_{t} \Gamma^{(2)}_{k}&=&\frac{1}{2}\kappa^{2}\int d^{4}x \sqrt{\bar{g}} \Big[-(\partial_{t}Z_{Nk})D^{2}-2\partial_{t}(Z_{Nk}\lambda_{k}) \nonumber \\
 & & +\partial_{t}Z_{Nk}C_{T}\bar{R}\Big] + 12 \frac{\partial_{t}(Z_{Nk}m_{k}^{2})}{2},
\end{eqnarray}
while for the terms in the right hand side we get

\begin{widetext}
\begin{eqnarray}
Tr[G_{k}\Gamma^{(3)}G_{k}\Gamma^{(3)}G_{k}\partial_{t}R_{k}]& = & Tr \Bigg[ \frac{\Big[43626 (D^{2})^{2} + 6504(-R+2\lambda_{k})(D^{2}) + 272(-R+2\lambda_{k})^{2}\Big]\partial_{t}R_{k}}{\Big[-D^{2}-2\lambda_{k} + k^{2}R(\frac{-D^{2}}{k^{2}}) + C_{R}R + \frac{m^{2}_{k}}{2}\Big]^{3}}\Bigg]  \nonumber \\
Tr [G_{k}\Gamma^{(4)}G_{k}\partial_{t}R_{k}] &=&   Tr\Bigg[\frac{ \Big[ - 100(D^{2}) -8(-R+2\lambda_{k})\Big]\partial_{t}R_{k}}{\Big[-D^{2}-2\lambda_{k} + k^{2}R(\frac{-D^{2}}{k^{2}}) + C_{R}R + \frac{m^{2}_{k}}{2}\Big]^{2}}\Bigg] \nonumber \\
\end{eqnarray}
\end{widetext}

Then we follow the same procedure described in section II, expanding the denominators and performing a Mellin transform on the $Tr[W(-D^{2})]$ expressions. By comparing the 
$\int d^{4}x \sqrt{\bar{g}}$ terms in both sides of (\ref{wetterich2deriv}) we obtain 

\begin{widetext}
 \begin{eqnarray}
-\partial_{t}(Z_{Nk}\lambda_{k})-3\partial_{t}(Z_{Nk}m_{k}^{2}) &=& \frac{10}{\kappa^{2}}(4\pi)^{-\frac{d}{2}}\Bigg[ 4\bar{\lambda}^{2}_{k}k^{d}\Big(2\cdot 272 \Big[2Q^{3}_{d/2} - \eta_{N}\widetilde{Q}^{3}_{d/2}\Big] - 3\cdot 272\frac{m^{2}_{k}}{2}\Big[2Q^{4}_{d/2} - \eta_{N}\widetilde{Q}^{4}_{d/2}\Big]\Big) \nonumber \\
& & + k^{d}8*2\bar{\lambda}_{k}\Big(\Big[2Q^{2}_{d/2} - \eta_{N}\widetilde{Q}^{2}_{d/2}\Big] - \frac{m^{2}_{k}}{2}\Big[2Q^{3}_{d/2} - \eta_{N}\widetilde{Q}^{3}_{d/2}\Big]\Big) \Bigg] 
\end{eqnarray}
\end{widetext}

This is a dimensionfull equation in the couplings. Combining this equation with the ones shown in (\ref{dimensionfulleqs}), we can obtain the mass beta function with the dimensionless couplings.
For the sharp cutoff this is given by 

\begin{widetext}
\begin{samepage}
\begin{eqnarray} 
 \partial_{t}m^{2}_{k}&=& -2m^{2}_{k}  + \frac{960 g_{k}}{\pi } \Bigg[\lambda^{2}_{k} \bigg(\frac{544}{(1-2 \lambda_{k})^2}-\frac{272 m^{2}_{k}}{(1-2 \lambda_{k})^3}\bigg)-\frac{m^{2}_{k}}{2 (1-2 \lambda_{k})^{2}}+\frac{16}{1-2 \lambda_{k}}\Bigg] \nonumber \\
& & +\frac{8 g_{k}}{\pi } \Bigg[20 \bigg(4 \zeta (3)-2\log (1-2 \lambda_{k}) + \frac{42 m^{2}_{k}}{1-2 \lambda_{k}} - 64 \zeta (3) \nonumber \\
& & -\frac{3g_{k}}{2  (5 g_{k} + 6 \pi )}\bigg(-\frac{2 m^{2}_{k}}{1-2 \lambda_{k}}-\frac{12}{1-2 \lambda_{k}}+\frac{10}{3} (\frac{\pi^{2}}{6}-\log (1-2 \lambda_{k})) -\frac{4 \pi^{2}}{9}-4\bigg)\bigg)\Bigg] \nonumber \\
& & +\frac{g_{k}m^{2}_{k}}{2 \pi } \Bigg[\frac{5}{3} \bigg( \frac{\pi ^2}{3}-2\log (1-2 \lambda_{k})-\frac{2 m^{2}_{k}}{1-2 \lambda_{k}}-\frac{12}{1-2 \lambda_{k}}-\frac{4 \pi ^2}{9}-1 \nonumber \\
& & -\frac{3g_{k}}{(5 g_{k} + 6 \pi )} \Big(-\frac{2 m^{2}_{k}}{1-2 \lambda_{k}}-\frac{12}{1-2 \lambda_{k}}+\frac{10}{3} (\frac{\pi ^2}{6}-\log (1-2 \lambda_{k}))-\frac{4 \pi ^2}{9}-4\Big)\bigg)\Bigg]
\end{eqnarray}
\end{samepage}
\end{widetext}

In this case we obtain an absolute non-gaussian fixed point(where $\beta_{m}=\beta_{\lambda}=\beta_{g}=0$) with coordinates $\lambda_{k}=0.3981,g_{k}=0.2192$ and $m_{k}=0.4221$ besides the IR fixed 
point for $\lambda_{k}=g_{k}=m_{k}=0$ where all the beta functions vanish.

As we can see, the original $m=0$ fixed point in the $\lambda_{k}-g_{k}$ plane gets slightly modified by the presence of a mass term.  
Notice that in this case, in which the mass is scale dependent, the limit $m^{2}_{k}\rightarrow0$ cannot be taken since the values of the critical exponents 
depend on the non gaussian fixed point, which is non zero for $m_k=0.4221$. This is an indication of the problems that are present in massive gravity. It seems 
that both approaches correspond to different physical theories. Nevertheless, looking at Fig. 1 it can be seen that for $m$ close to $0.4221$, which is the value 
that gives our second approach for the fixed point mass, the values of $\lambda_{k}$ and $g_{k}$ are approximately those that are obtained in the second approach. 
This shows the consistency between both calculations.

For the critical exponents in the UV fixed point, we obtain the values $\theta_{1}=342408$, $\theta_{2}=7.91$ and $\theta^{*}_{3}=5.09$, 
while for the gaussian fixed point ($g_{k}=\lambda_{k}=m_{k}=0$) we obtain the critical exponents $\theta_{1}=\theta_{2}=2$ and $\theta_{3}=-2$.
Clearly, these values show the difficulties of massive gravity, which we certainly expected to encounter, since all approaches to massive gravity at a certain point produce results that have no physical significance. It could also be
that there are inconsistencies in the truncation, which means that we need to consider non-linear terms (interactions). Another possibility could be that, since we are working in a theory  that
is not manifestly gauge invariant, which can be stated as a particular gauge choice, we could restore gauge invariance
through the Stueckelberg mechanism, and analyze this more general theory.

\section{Comments and Conclusions}

We have analyzed massive gravity in its simplest version (Fierz-Pauli), within the Asymptotic Safety renormalization group formalism. In a first approach, we considered the mass to be 
a fixed parameter, and found that we get different fixed points for the other scale dependent coupling constants (Newton and cosmological constant) as we change this mass parameter.
In this case there is no discontinuity in the $m\rightarrow0$ limit, so there is no vDVZ discontinuity and the massless limit appears as a continuous limit
of the massive case and the flow diagram changes continuously as $m^{2}_{k}$ grows. This may be because in order to obtain the threshold functions an expansion is usually made 
around the Ricci scalar and in this case the expansion was made around the term $R + c_{m}m^{2}_{k}$ with $c_{m}$ constant, so the mass term acts as a correction to the curvature.

The result that with this procedure there is no vDVZ discontinuity is important, because at present the only solution to this problem is given by the Stueckelberg mechanism, 
and also because the variation in the mass parameter affects the value of the cosmological and gravitational constants, with no discontinuity.

The mass term, which is an infrared modification of gravity, changes the coordinates of the non gaussian fixed point. This means that an infrared modification still produces
an effect in the UV sector of the theory. On the other hand, the gaussian fixed point, $\lambda_{k}=g_{k}=0$, does not change with the mass term. 

In the case where the mass term is taken as a constant, the critical exponents (fig.\ref{criticalexponents}) deviate from the ``expected'' values ($\sim 2$). This fact is 
consistent with the upper bound of the graviton mass predicted by experimental measurements, which gives a maximum value of order $\sim 10^{-23}$ eV (smaller than the upper limit 
of the photon mass  $< 10^{-18}$ eV). Therefore, if the graviton mass is not equal to zero, it must be small so the truncation gives consistent results.

Nevertheless, we should stress that there is a consistency check that can be made between both cases. In our first approach with a constant mass, looking at Fig. 1 it can be seen that for $m$ close to $0.4221$, which 
is the value that gives our second approach for the fixed point mass, the values of $\lambda_{k}$ and $g_{k}$ are approximately those that are obtained in the second approach. This shows the 
consistency between both calculations.

On the other hand, for the case with the mass as a scale dependent parameter, we still obtain a non-gaussian fixed point as Asymptotic Safety demands, but the values of the critical 
exponents indicate that there are serious problems. In our view, these values show the expected difficulties of massive gravity, that are present in general in all approaches to massive gravity, 
which produce results that have no physical significance. This could also be related to inconsistencies in the truncation, which means that we need to consider non-linear terms (interactions). 
These have been shown to have important effects in the classical theory \cite{Vainshtein}, and come from two sources. First we have those non-linear terms that arise just from General Relativity,
and also those that are present in an extra interaction potential. The first case leads to difficulties such as ghost instabilities, while the last one has a lot of freedom (choices of parameters).
One version, named $\Lambda_{3}$, seems to solve some of the problems, and is therefore the best present possibility of having a sound massive gravity theory. Another possibility is that, since we 
are working in a theory that is not manifestly gauge invariant, we could use the Stueckelberg mechanism in order to promote this theory into a gauge invariant version, which could clarify the origin on the unphysical results. 
This is an approach that has worked in similar instances, such as the clarification of the vDVZ discontinuity \cite{Hinterbichler}.  It is certainly important to consider these extensions from the point of view of the renormalization group and asymptotic safety,
which is something that we tend to do in future publications.

\begin{acknowledgments}

We would like to thanks O. Castillo for his collaboration in the preparation of this manuscript. M.B. is also grateful to Arkady Vainshtein for hospitality and useful discussions
during his stay at University of Minnesota.
This work is supported by $MECESUP_{2}$(Chile) under grant FSM0806-D3042, DGIP-UTFSM(Chile), UTFSM-PIIC project, Fondecyt (Chile) grant 1180232 and by ANID PIA/APOYO AFB180002.   

.

\end{acknowledgments}

\appendix

\section{Denominator Expansion}\label{sss}

In order to obtain the right hand side for the evolution equation (\ref{evolutionequation}), after expanding the metric in its background and fluctuating part we obtain, two terms of the form

\begin{equation}
 (\Gamma^{(2)} + R_{k})^{-1}_{\hat{\Phi}\hat{\Phi}}\partial_{t}R_{k} = \frac{N}{A+C_{i}R+C_{m}m^{2}_{k}}
\end{equation}
with $A=-D^{2}-2\lambda_{k}$, $N=(2-\eta_{N}(k))k^{2}R_{k}(\frac{-D^{2}}{k^{2}}) + 2 D^{2}R_{k}'(\frac{-D^{2}}{k^{2}})$ and $C_{i}$, and $C_{m}$ two constants which depend on the dimension $d$
 and the field $\Phi=(\hat{h},\phi)$. The term $(\Gamma^{2}+R_{k})$ is known as the inverse modified propagator \cite{QEG}. In order to compare the right and left hand side of (\ref{evolutionequation}) using the heat kernel expansion we need the R-term to be on the numerator of 
these expressions, so an expansion is made around $C_{i}R$ acording to $N(A+C_{s}R)^{-1}=NA^{-1}+NA^{-2}C_{s}R$.

\begin{widetext}
\begin{figure}
\begin{center}
\begin{tabular}{ccc}
	  \includegraphics[width=0.3\textwidth]{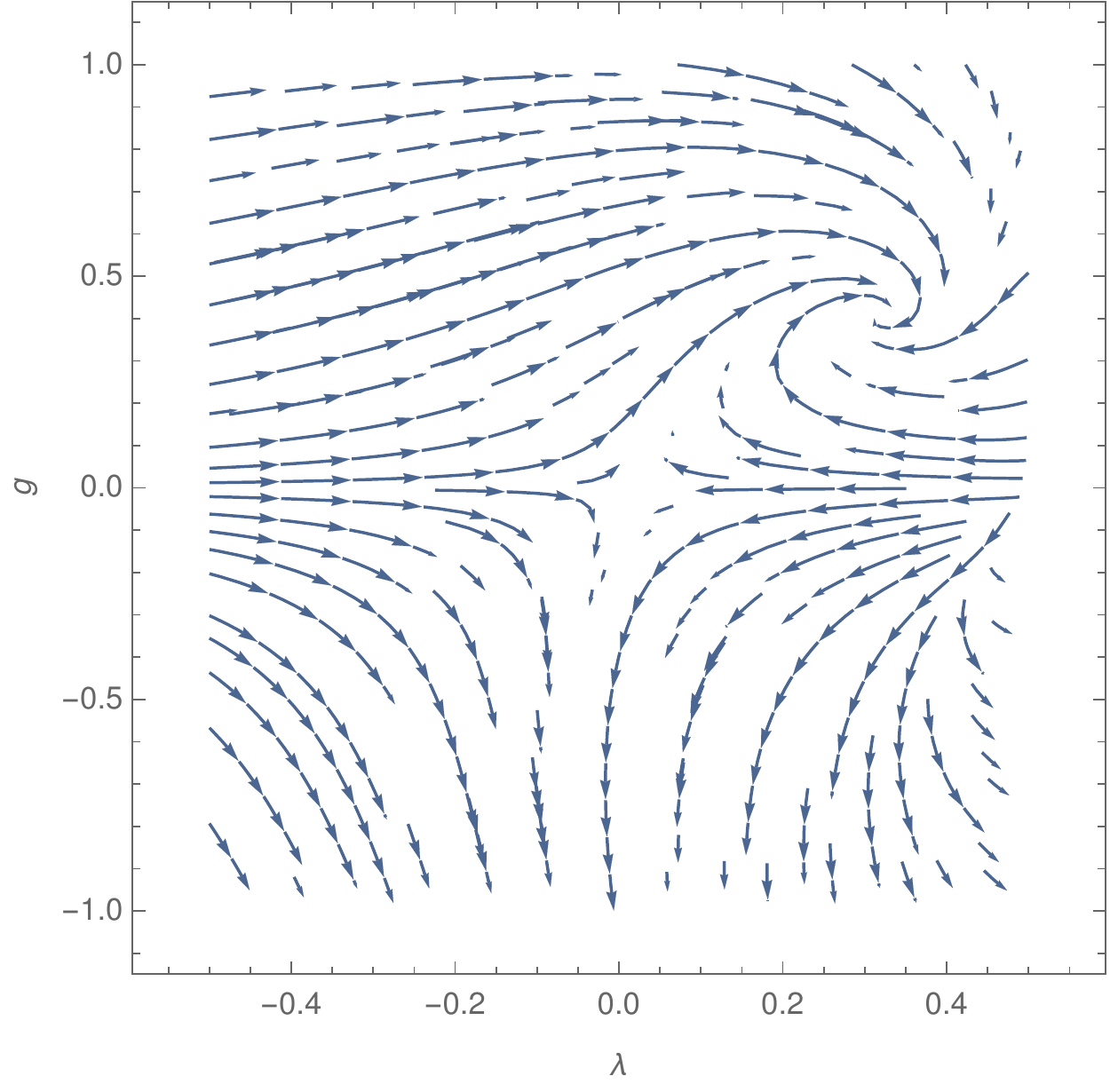}
&	  \includegraphics[width=0.3\textwidth]{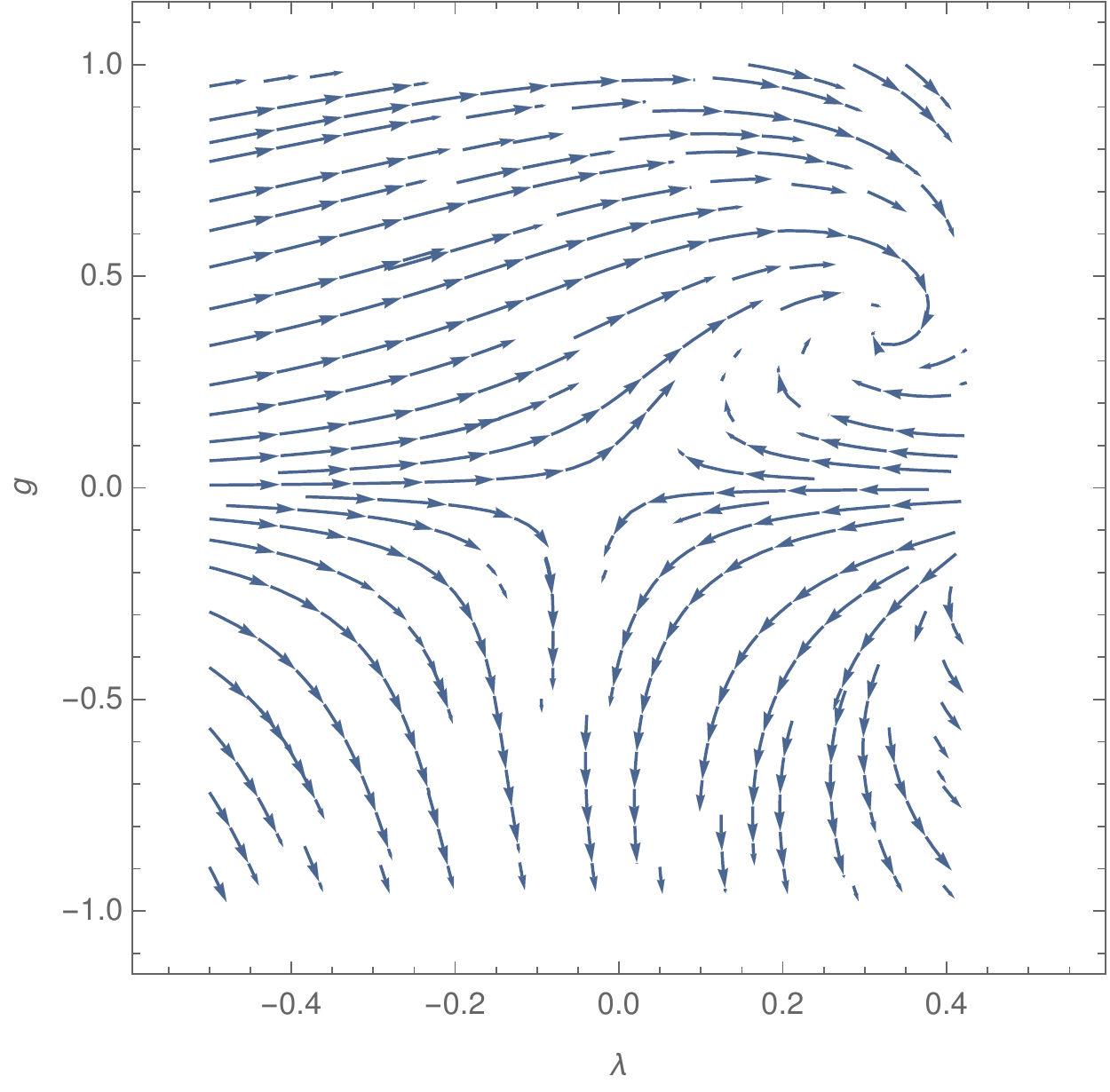}
&	  \includegraphics[width=0.3\textwidth]{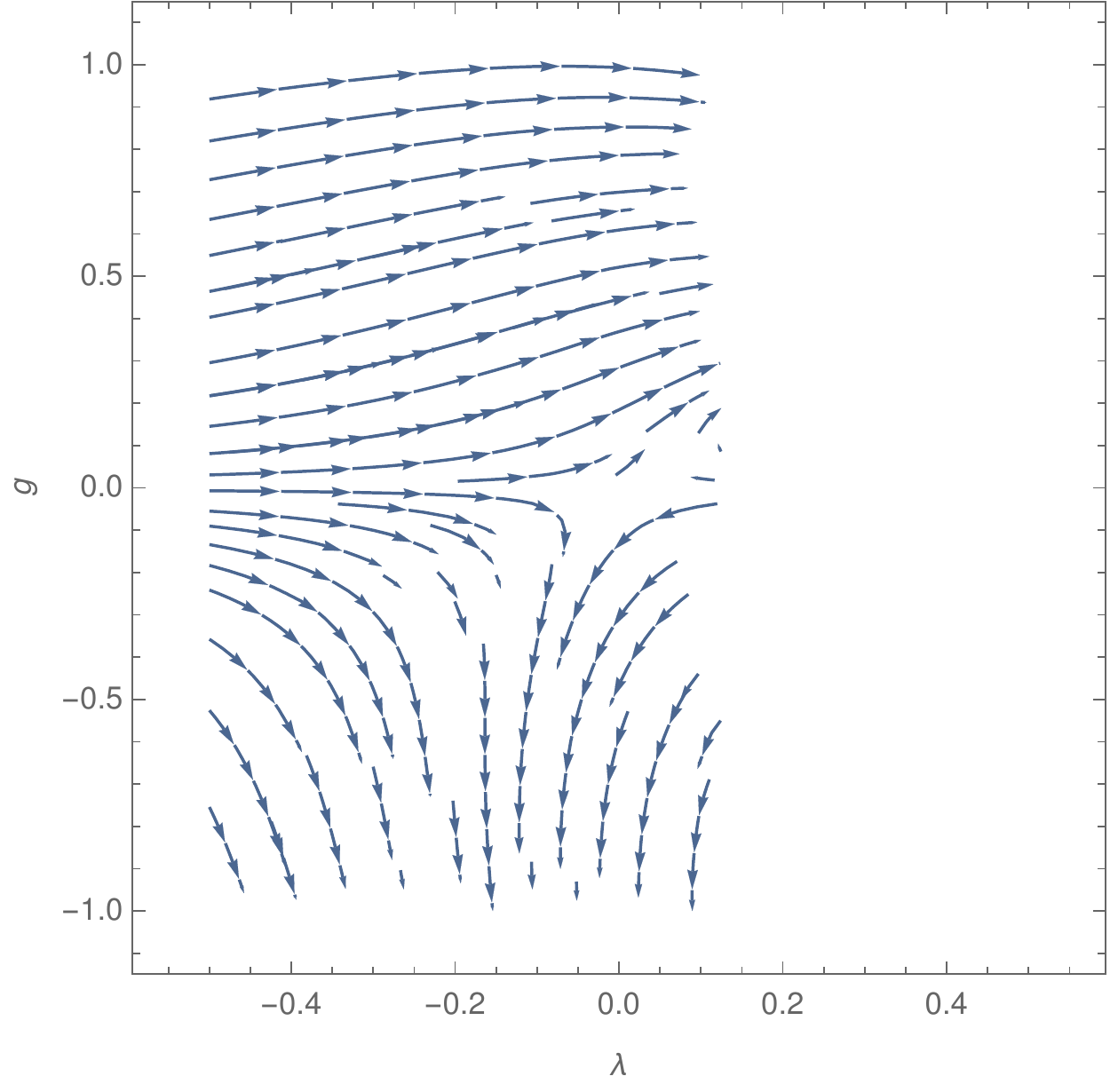}
\end{tabular}
\caption{Flow diagrams for the expansion made around $C_{i}R$ for $m^{2}_{k}=0,0.1$ and $0.5$. Here $\lambda_{k}$ ($g_{k}$) is in the x (y) axis.}
\label{flujodistintasmasasR}
\end{center}
\end{figure}
\end{widetext}.

If we perform the expansion around $C_{i}R$ (as is done in \cite{Reuter98}), we obtain terms in the beta functions of the form $log(1-2\lambda_{k}-C_{m}m^{2}_{k})$ instead of $log(1-2\lambda_{k})$ and the flow diagrams are the ones shown in fig(\ref{flujodistintasmasasR}).

As we can see, the argument in $log(1-2\lambda_{k}-C_{m}m^{2}_{k})$ becomes negative and we can no longer obtain information from the beta functions. To avoid this problem, we perform the same expansion but around $C_{i}R + C_{m}m^{2}_{k}$ to obtain the flow equations in (\ref{funcionesbeta}). It is worth mentioning that both approaches 
give the same flow of the couplings for the $m\rightarrow 0$ limit.

\end{document}